\documentclass[10pt,journal]{IEEEtran}

\usepackage{outline}
\usepackage{pmgraph}
\usepackage[normalem]{ulem}
\usepackage[utf8]{inputenc}
\usepackage{amssymb}
\usepackage{hyperref}
\usepackage{amsmath}
\usepackage{graphicx}
\usepackage{times}
\usepackage{xcolor}
\usepackage{xspace}
\usepackage[colorinlistoftodos]{todonotes} 
\usepackage{cite}
\usepackage{bm} 
\usepackage{url}

\usepackage{epstopdf}
\AppendGraphicsExtensions{.tiff}
\graphicspath{{fig/}} 

\usepackage{epsfig}
\usepackage{tikz}
\usetikzlibrary{spy}
\usepackage{algpseudocode}
\usepackage{algorithm}
\usepackage{mathrsfs}

\def\QED{~\rule[-1pt]{5pt}{5pt}\par\medskip}

\long\def\comment#1{} 

\newcommand{\Xc}{\mathcal{X}}
\newcommand{\Yc}{\mathcal{Y}}

\newcommand{\Rd}{{\mathbb R}}

\newcommand{\Zc}{{{\mathcal Z}}}

\newcommand{\beq}{\begin{equation}}
\newcommand{\eeq}{\end{equation}}
\newcommand{\beqa}{\begin{eqnarray}}
\newcommand{\eeqa}{\end{eqnarray}}

\def\BibTeX{{\rm B\kern-.05em{\sc i\kern-.025em b}\kern-.08emT\kern-.1667em\lower.7ex\hbox{E}\kern-.125emX}}

\usepackage{cite}
\usepackage{bm} 
\usepackage{hyperref}
\usepackage{xcolor}

\begin{document}

\title{Unsupervised CT Metal Artifact Learning using Attention-guided $\beta$-CycleGAN}

\author{Junghyun Lee, Jawook Gu, and Jong Chul Ye, \IEEEmembership{Fellow, IEEE}
\thanks{
Authors are with the Department of Bio and Brain Engineering, Korea Advanced Institute of Science and Technology (KAIST), Daejeon 34141, Republic of Korea (E-mail: \{jh200.lee,jwisdom9299,jong.ye\}@kaist.ac.kr).}}
\maketitle
\begin{abstract}
Metal artifact reduction (MAR) is one of the most important research topics in computed tomography (CT). 
With the advance of deep learning technology for image reconstruction,
various deep learning methods have been also suggested for metal artifact removal, among  which supervised learning methods are most popular. 
However, matched non-metal and metal image pairs  are difficult to obtain in real CT acquisition.
Recently, a promising unsupervised learning for MAR  was proposed using feature disentanglement, but the resulting network architecture
is complication and difficult to handle large size clinical images.
To address this, here we propose a much simpler and much effective
   unsupervised MAR method for CT.
  The proposed method is based on a novel $\beta$-cycleGAN architecture derived
  from the optimal transport theory for appropriate feature space disentanglement. 
Another important contribution is to show that attention mechanism is the key element to effectively remove the metal artifacts.
Specifically,  by adding the convolutional block attention module (CBAM) layers with a proper disentanglement parameter, experimental results confirm that
we can get more improved MAR  that preserves the detailed texture of the original image.
\end{abstract}
\section{Introduction}
\label{sec:introduction}
\IEEEPARstart{X}{-ray} computed tomography (CT) is  widely
used  for dental applications in recent years. 
Most commercially available dental CT scanners  reconstruct the stack of transversal images of the jaw, which is scanned parallel to the alveolar ridge of the teeth using a rotating x-ray source and a flat panel detector. For a circular source trajectory, an approximated inversion algorithm, called the Feldkamp, Davis and Kress (FDK) algorithm\cite{feldkamp1984practical}, is the most widely used.  Even though the FDK algorithm introduces conebeam artifacts, the amount of them in dental applications is minimal since the region of interest is usually the area around the jaw at the mid-plane of the circular trajectory. On the other hand, the more serious problem in dental CT is  the typical placements of metallic implants and dental fillings that can cause severe metal artifacts.
Similar metal artifacts are quite often encountered in musculoskeletal CT imaging of the patients with metallic implants. 

\begin{figure}[!tb]
\centering
{\includegraphics[width=0.8\columnwidth]{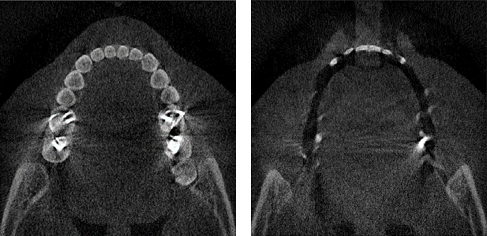}}
\caption{Examples of dental CT images with  metal artifacts.}
\vspace*{-0.5cm}
\label{fig_metal_artifact_samples}
\end{figure}

 In the imaging of patients with metallic inserts, X-ray photons cannot penetrate the metallic object consistently due to the object's high attenuation. This causes severe streaking and shading artifacts that deteriorate the image quality in reconstructed images as shown in Fig.~\ref{fig_metal_artifact_samples}. Other reasons such as beam-hardening or poor signal to noise ratio (SNR) can contribute to metal artifacts\cite{tomography2003principles,zhao2000x}.

There are some traditional  methods that modify the sinogram and reconstruct objects by removing the corrupted sinogram and interpolating it from adjacent data\cite{zhao2000x,kalender1987reduction,mahnken2003new,meyer2010normalized}. But these methods have a limitation for general applications due to the difficulty of optimal parameter selection. Iterative reconstuction is another method of metal artifact removal, which includes 
expectation maximization\cite{shepp1982maximum,wang1996iterative} and iterative maximum-likelihood polychromatic algorithm for CT\cite{de2001iterative}. But their main drawback is the extremely high computational complexity.

Recently, motivated by the success of deep learning, several works using deep learning for MAR have been proposed \cite{wang2018conditional,zhang2018convolutional,lin2019dudonet}. The advantage  of using deep learning methods comes from the data-driven nature that automatically learns the optimal features for the task from  the data. There are some examples for MAR,  which applied the pix2pix model\cite{wang2018conditional}, which proposed to first estimate a prior image by a convolutional neural network (CNN)\cite{zhang2018convolutional}, and which proposed sinogram network and the image network by learning two CNNs\cite{lin2019dudonet}. But these networks are trained in a  supervised manner. Accordingly, they require structurally matched images including the metal artifacts and the clean target images in pairs, which are hard to get in real situation.  Although one could use simulation data to train the neural network,
 due to the complexity of metal artifacts and the variations of CT devices, the synthesized images may not fully reflect the real clinical scenarios so that the performances of these supervised methods may degrade in clinical applications.

To utilize pairs of unmatched images, unsupervised learning approaches should be used. Among the various approaches for unsupervised learning, generative adversarial network (GAN)\cite{goodfellow2014generative} can learn how to match a distribution of the input domain to a distribution of the target domain. However, this approach often suffers from the mode-collapse behavior, which often generates artificial features. To address the mode-collapse problem, an unsupervised image-to-image translation task using the cycle-consistent adversarial network (CycleGAN) was proposed\cite{zhu2017unpaired}. Specifically, the network is trained in an unsupervised manner using generative networks, and the cyclic consistency alleviates the generation of artificial features due to the mode collapsing problem of GAN.  
Recently, the mathematical origin of cycleGAN was revealed using optimal transport theory \cite{villani2008optimal,peyre2019computational} as an 
unsupervised distribution matching between two probability spaces \cite{sim2019optimal}.
Therefore, we are  interested in utilizing the cycleGAN as an interpretable backbone for our unsupervised metal artifact removal.
Another important advantage of cycleGAN is that once the neural network is trained, only a single generator is necessary at the inference
stage, which makes the algorithm simple.

Because the metal artifacts occur both local and global patterns as shown Fig. \ref{fig_metal_artifact_samples}, it is hard to train by small patches. 
On the other hand, the metal artifacts have unique characteristics in that they are radiated from a few metalic regions.
Therefore, we need a simple yet effective module that can boost representation power of the network by focusing only where the artifacts exist and how their patterns appear. To address these issues, a method mimicking the human visual system can be a good option because 
humans exploit a sequence of partial glimpses and selectively focus on salient parts in order to capture the visual structure much better\cite{larochellesupplementary}. This  attention mechanism  in human visual system\cite{itti1998model,rensink2000dynamic,corbetta2002control} has inspired  recent
advance of attention modules in deep neural networks.
Among the modules applying attention mechanism, the convolutional block attention module (CBAM)\cite{woo2018cbam} is one of the simplest yet effective one. 
Accordingly,  we  propose an attention-guided unsupervised MAR method using the cycle-consistent adversarial network with CBAM.

Yet another important contribution of this work is the introduction of the parameters to control the level of feature disentanglement.
Specifically, inspired by the success of $\beta$-VAE (variational auto-encoder) \cite{higgins2017beta},  we control the level
of the importance in terms of the statistical distances in the original and target domains using a weighting parameter $\beta$.
It turns out that $\beta$-parameter plays an important role in MAR for real data by emphasizing the faithfulness of the recovered
images. 

\section{Related Works}
\subsection{Conventional methods}
Among the classical MAR algorithms, the sinogram modification methods reconstruct objects after the corrupted sinogram is removed and interpolated from adjacent data. For example, linear interpolation (LI) is the most traditional and simplest method\cite{kalender1987reduction}. It replaces the metallic parts in the original sinogram with linear interpolated values from the boundaries. Although LI removes most background artifacts, this usually causes new artifacts due to inaccurate values interpolated in the metallic parts in the sinogram. Other interpolation methods have been suggested to improve LI\cite{zhao2000x,kalender1987reduction,mahnken2003new,meyer2010normalized}, among which
normalized metal artifact reduction (NMAR) \cite{meyer2010normalized} is most well-known. These methods improve image quality, but they still have a limitation for general applications due to the difficulty of optimal parameter selection. 

Other works use iterative reconstruction methods for MAR such Maximum-Likelihood for TRansmission (ML-TR)\cite{de2000reduction}, expectation maximization (EM)\cite{shepp1982maximum,wang1996iterative} and iterative maximum-likelihood polychromatic algorithm for CT (IMPACT)\cite{de2001iterative}. The main idea of those iterative approaches is to take sinogram inconsistency into consideration by correctly modeling its physical origin. While the reconstruction results from iterative reconstruction are nearly free of metallic artifacts, one of their main drawbacks is extremely high computation complexity.

\subsection{Unsupervised MAR models}
To deal with the limitation of supervised deep learning methods for MAR as reviewed in the introduction,
unsupervised  MAR has been recently proposed \cite{liao2019adn,ranzini2020combining}. In particular, Liao et al.\cite{liao2019adn} proposed an artifact disentanglement network for the unsupervised metal artifact reduction (ADN).
The ADN method disentangles the artifact and content components of an artifact-affected image by encoding them separately into a content space and an artifact space. If the disentanglement is well addressed, the encoded content component should contain no information about the artifact while preserving all the content information. {However, the network architecture for  ADN network is highly complicated
due to the explicit disentanglement steps. Moreover, as will be shown later, due to this explicit disentanglement
using artifact-free images,
the ADN often introduces artificial features when the input image is corrupted with other artifacts that have not been considered
during the training. Therefore, we found that ADN fails to produce
any meaningful MAR for dental CT applications.} 
 Moreover, the method proposed by Ranzini et al.\cite{ranzini2020combining}, which is similar to ADN, uses paired MRI and CT images. For dental CT, it is too costly to get paired images in practice and difficult to apply the method accordingly.



\subsection{Attention model} 
Many generative adversarial network (GAN)-based on the deep convolutional networks had difficulty in modeling some image classes more than others when training on multi-class datasets \cite{odena2017conditional,lucas2018mixed}. They failed to capture the geometric or structural patterns that occur consistently in some classes. This is because small receptive field from convolution operator may not be able to represent them, optimization algorithms may have trouble discovering parameter values that carefully coordinate multiple layers to capture these dependencies, and these parameterizations may be brittle and prone to failure when applied to previously unseen inputs. Increasing the size of the convolution kernels can increase the representational capacity of the network but doing so also could lose the computational and the statistical efficiency obtained by CNN.

As such, attention mechanisms have become an integral part of models that must capture global dependencies, since attention
is designed to capture global patterns. There are some models  using a self attention mechanism  that calculates the response at a position in a sequence by attending to all positions within the same sequence. Among those models, the self attention Generative Adversarial Network (SAGAN)\cite{zhang2018self} is the popular one that uses self attention in the context of GAN. SAGAN efficiently learns to find global and long-range of dependencies within internal representations of images. Using two matrix multiplication operations to compute
the key and query, it is able to make the model to be effective in obtaining information about the entire spatial regions. Unfortunately,
calculating the key and query from entire images is often computationally expensive and can cause memory problems as the spatial sizes of input get bigger. 

Therefore, we used the CBAM module \cite{woo2018cbam}, which is simple yet effective to obtain information from wide regions compared to other attention modules. Details on CBAM will be discussed in the following theory section.

\subsection{$\beta$-VAE for feature space disentanglement} 

The idea of Variational Autoencoder \cite{kingma2013auto}, short for VAE, is  deeply rooted in the methods of variational Bayesian and graphical model.
Specifically,  in VAE,  a given data set $x\in \Xc$ is 
modeled as a parameterized distribution $p_\theta(x)$, 
and the goal is to find the parameter $\theta$ to maximize the loglikelihood.
As the direct modeling of $p_\theta(x)$ is difficult,  it is modeled  by combining a simple distribution 
$p(z),z\in \Zc$ in the latent space $\Zc$  with a  family of conditional distributions $p_\theta(x|z)$, so that our objective is written
by
\begin{align}
\log p_\theta(x) 
&= \log\left(\int p_\theta(x|z)p(z) dz \right)\notag\\
&=\log\left(\int p_\theta(x|z)\frac{p(z)}{q_\phi(z|x)}q_\phi(z|x) dz \right)
\end{align}
where  $p_\theta(x|z)$ is the conditional probability for a given $z$,
 $q_\phi(z|x)$ is a user-chosen posterior distribution model parameterized by $\phi$.
Using the Jensen's inequality, this leads to the well-known evidence-lower bound (ELBO) loss function as an upper-bound
of $-\log p_\theta(x)$ \cite{kingma2013auto}: 
\begin{align}
&\ell_{ELBO}(x;\theta,\phi)\notag\\
&:=-\int \log p_\theta(x|z) q_\phi(z|x)dz +D_{KL}(q_\phi(z|x)||p(z))   \label{eq:ELBOloss}
\end{align}
where $D_{KL}$ denotes the Kullback-Leiber (KL) divergence \cite{kullback1997information}.

By inspection of VAE loss in \eqref{eq:ELBOloss}, we can easily see that
the first term represents the distance between the generative samples and the real ones,
whereas the second term is the KL distance between the real latent space measure
and posterior distribution.
Therefore, VAE loss is a measure of the distances that equally considers both latent  space and the ambient
space between real and generated samples.

Rather than giving uniform weights for both distances,
$\beta$-VAE \cite{higgins2017beta}  introduces a controllable parameter $\beta$ to impose the relative importance
between the two distances:
\begin{align}
&\ell_{\beta-VAE}(x;\theta,\phi)\notag\\
&:=-\int \log p_\theta(x|z) q_\phi(z|x)dz + \beta D_{KL}(q_\phi(z|x)||p(z))   \label{eq:betaloss}
\end{align}
As a higher $\beta$  imposes more constraint on the latent space, it turns out that the latent
space is more interpretable and controllable, which is known as the {\em disentanglement}.
One benefit that often comes with disentangled representation is that it is only sensitive to one single generative factor and relatively invariant to other factors, 
leading to good interpretability and easy generalization to a variety of tasks.
On other other hand, large $\beta$ values give  less emphasis on the reproduction quality,
which produces more blurry results  than those with $\beta=1$.

\section{Theory}


\subsection{Geometry of CycleGAN}

CycleGAN has shown great performance especially in unsupervised image artifact removal.
Kang et al.\cite{kang2018deep}  proposed a CycleGAN-based-model for the removal of Low-Dose CT noise,
and Song et al.\cite{song2020unsupervised} also proposed a CycleGAN-based-model for the removal of noise in satellite imagery. 
Given the success,  one is interested whether the resulting improvement is real or cosmetic changes.

\begin{figure}[!hbt] 	
\center{ 
\includegraphics[width=8cm]{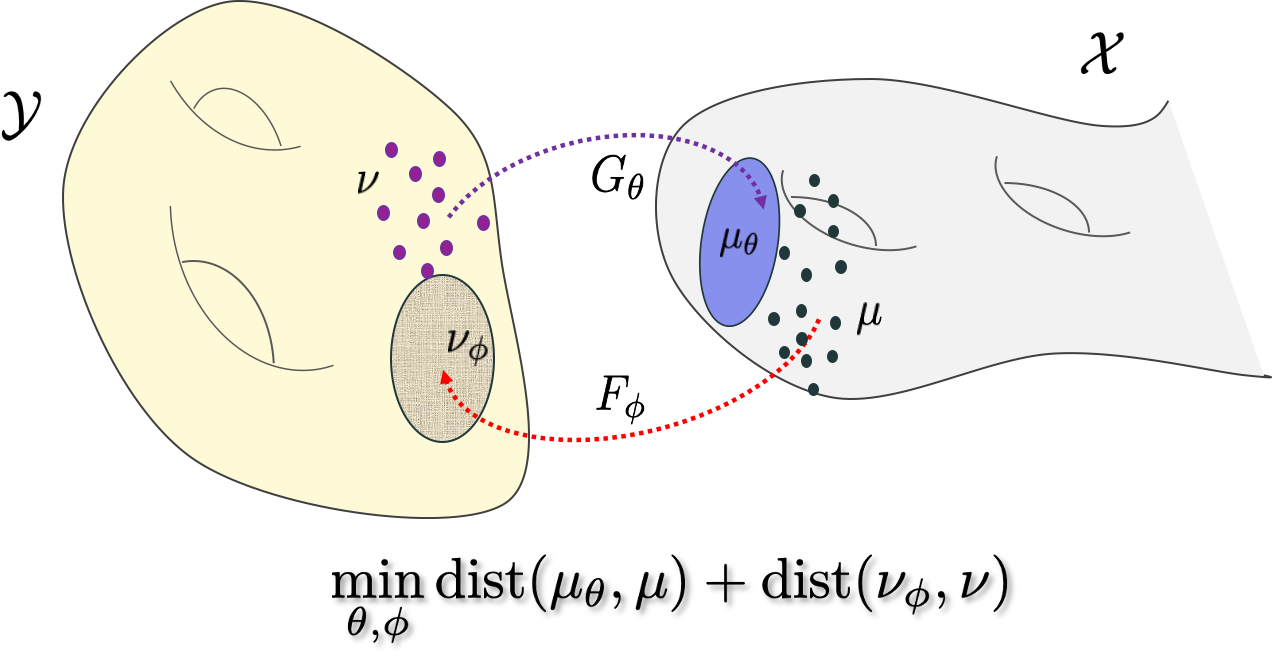}
}
\vspace*{-0.5cm}
\caption{Geometric view of unsupervised learning.}
\label{fig:cycleGANgeom}
\end{figure}

In that sense,  optimal transport (OT)  \cite{villani2008optimal,peyre2019computational} 
 provides a rigorous mathematical tool to understand the geometry of unsupervised learning by cycleGAN.
Our geometric
view of unsupervised learning  from optimal transport theory is shown in Fig.~\ref{fig:cycleGANgeom}.
Here, the target image space
$\Xc$ is equipped with a probability measure $\mu$, whereas
the original image space  $\Yc$ is  with a probability measure $\nu$.
Since there are no paired data, the goal of unsupervised learning is to match the probability distributions rather than each individual samples.
This can be done by finding  transportation maps that transport the measure $\mu$ to $\nu$, and vice versa.

More specifically, the transportation from a measure space $(\Yc,\nu)$ to another measure space $(\Xc,\mu)$ is done by a generator $G_\theta: \Yc \mapsto \Xc$, realized by a deep
network parameterized with $\theta$.  Then, the generator $G_\theta$ ``pushes forward'' the measure $\nu$ in $\Yc$ to a measure $\mu_\theta$ in the target space $\Xc$ \cite{villani2008optimal,peyre2019computational}. Similarly, the transport from $(\Xc,\mu)$ to $(\Yc,\nu)$ is performed by another neural network generator $F_\phi$,
 so that the generator $F_\phi$ pushes forward the measure $\mu$ in $\Xc$ to $\nu_\phi$ in the original space $\Yc$.
Then, the optimal transport map for unsupervised learning can be achieved by minimizing the statistical distances  $\mathrm{dist}(\mu_\theta,\mu)$ between $\mu$ and $\mu_\theta$, and  $\mathrm{dist}(\nu_\phi,\nu)$ 
 between $\nu$ and $\nu_\phi$, and our proposal is to use the Wasserstein-1 metric as a means to measure the statistical distance.

More specifically, for the choice of a metric $d(x,x')=\|x-x'\|$ in $\Xc$,  
the Wasserstein-1 metric between $\mu$ and $\mu_\theta$ can
be computed by \cite{villani2008optimal,peyre2019computational}
\begin{align}\label{eq:Wmu}
W_1(\mu,\mu_\theta)
=&\inf\limits_{\pi \in \Pi(\mu,\nu)}\int_{\Xc\times \Yc} \|x-G_\theta(y)\|d\pi(x,y) 
\end{align}
Similarly, the Wasserstein-1 distance between $\nu$ and $\nu_\phi$ is given by
\begin{align}\label{eq:Wnu}
W_1(\nu,\nu_\phi)
=&\inf\limits_{\pi \in \Pi(\mu,\nu)}\int_{\Xc\times \Yc} \|F_\phi(x)-y\|d\pi(x,y) 
\end{align}
Rather than minimizing \eqref{eq:Wmu} and \eqref{eq:Wnu} separately with distinct joint distributions,
a better way of finding the transportation map is to minimize them together with the same joint distribution $\pi$:
\begin{align}\label{eq:unsupervised}
\inf\limits_{\pi \in \Pi(\mu,\nu)}\int_{\Xc\times \Yc} \|x-G_\theta(y)\|+ \|F_\phi(x)-y\|d\pi(x,y) 
\end{align}
One of the most important contributions of our companion paper \cite{sim2019optimal} is to show that
the primal formulation of the unsupervised learning in \eqref{eq:unsupervised} 
can be represented by a dual formulation:
\begin{eqnarray}\label{eq:OTcycleGAN}
\min_{\phi,\theta}\max_{\psi,\varphi}\ell_{cycleGAN}(\theta,\phi;\psi,\varphi)
\end{eqnarray}
where 
\begin{eqnarray}
\ell_{cycleGAN}(\theta,\phi;\psi,\varphi):=  \lambda \ell_{cycle}(\theta,\phi) +\ell_{Disc}(\theta,\phi;\psi,\varphi) 
\end{eqnarray}
where $\lambda>0$ is the hyper-parameter, and  the cycle-consistency term is given by
\begin{align*}
\ell_{cycle}(\theta,\phi)  =& \int_{\Xc} \|x- G_\theta(F_\phi(x)) \|  d\mu(x) \\
&+\int_{\Yc} \|y-F_\phi(G_\theta(y))\|   d\nu(y)
\end{align*}
whereas  the second term is
\begin{align}
&\ell_{Disc}(\theta,\phi;\psi,\varphi)  \label{eq:disc} \\
=&\max_{\varphi}\int_\Xc D_\varphi(x)  d\mu(x) - \int_\Yc D_\varphi(G_\theta(y))d\nu(y) \notag \\
 & + \max_{\psi}\int_{\Yc} D_\psi(y)  d\nu(y) - \int_\Xc D_\psi(F_\phi(x))  d\mu(x) \notag
\end{align}
Here, $\varphi,\psi$ are often called Kantorovich potentials and satisfy 1-Lipschitz condition (i.e.
\begin{align*}
|D_\varphi(x)-D_\varphi(x')|\leq \|x-x'\|,&~\forall x,x'\in \Xc \\
|D_\psi(y)-D_\psi(y')|\leq \|y-y'\|,&~\forall y,y'\in \Yc
\end{align*}
In machine learning context,  the 1-Lipschitz potentials  $\varphi$ and $\psi$
correspond to the  Wasserstein-GAN (W-GAN) discriminators \cite{arjovsky2017wasserstein}.
Specifically, 
$\varphi$ tries to find the difference between the true image $x$ and the generated image $G_\Theta(y)$,
whereas $\psi$ attempts to find the fake measurement data  that are generated by the synthetic
measurement procedure $F_\phi(x)$.
In fact, this formulation is equivalent to the cycleGAN formulation \cite{zhu2017unpaired} except for the use of 1-Lipschitz discriminators.
%
In our companion paper \cite{lim2019cyclegan}, we further showed that the popular LS-GAN approach \cite{mao2017least}, which is often used in combination of standard
cycleGAN \cite{zhu2017unpaired},  is also closely related to
imposing the finite Lipschitz condition.
In this paper, we  therefore consider LS-GAN variation as our discriminator term.

\subsection{$\beta$-CycleGAN for metal artifact disentanglement}

In the application of cycleGAN for MAR, we assume that $\Yc$ is the domain for metal-artifact images,
whereas $\Xc$ is the artifact-free images.  Inspired by the success of $\beta$-VAE\cite{higgins2017beta},
our goal is to give an unequal weight
on the statistical distances in  $\Xc$ and  $\Yc$.
This is done using the following loss function:
\begin{align}\label{eq:betaprimal}
\inf\limits_{\pi \in \Pi(\mu,\nu)}\int_{\Xc\times \Yc} \|x-G_\theta(y)\|+ \frac{1}{\beta}\|F_\phi(x)-y\|d\pi(x,y) 
\end{align}
where we use the reciprocal weighting to the statistical distance in $\Yc$ for  notational simplicity in this paper.

As shown in Appendix, the corresponding dual loss function for the primal problem in \eqref{eq:betaprimal} is given by
\begin{eqnarray}\label{eq:Dd}
\ell_{\beta-cycleGAN}(\theta,\phi;\psi,\varphi):=  \lambda \ell_{\beta-cycle}(\theta,\phi) +\ell_{Disc}(\theta,\phi;\psi,\varphi)
\end{eqnarray}
where $\ell_{Disc}$ is the same as in \eqref{eq:disc}, whereas the cycle-consistency term becomes
\begin{align}
\ell_{\beta-cycle}(\theta,\phi)  =& \int_{\Xc} \|x- G_\theta(F_\phi(x)) \|  d\mu(x) \notag \\
&+\frac{1}{\beta}\int_{\Yc} \|y-F_\phi(G_\theta(y))\|   d\nu(y) \label{eq:betacycle}
\end{align}
Another mathematical difference is that the discriminators $D_\varphi$ and $D_\psi$ are now $1/\beta$-Lipschitz as shown in Appendix. 
However, this does not cause any practical changes in the discriminator implementation, since most of the regularization for discriminator
\cite{mao2017least,gulrajani2017improved} are invariant with respect to the Lipschitz constant variation.

\begin{figure}[!htb]
\centerline{\includegraphics[width=1.0\columnwidth]{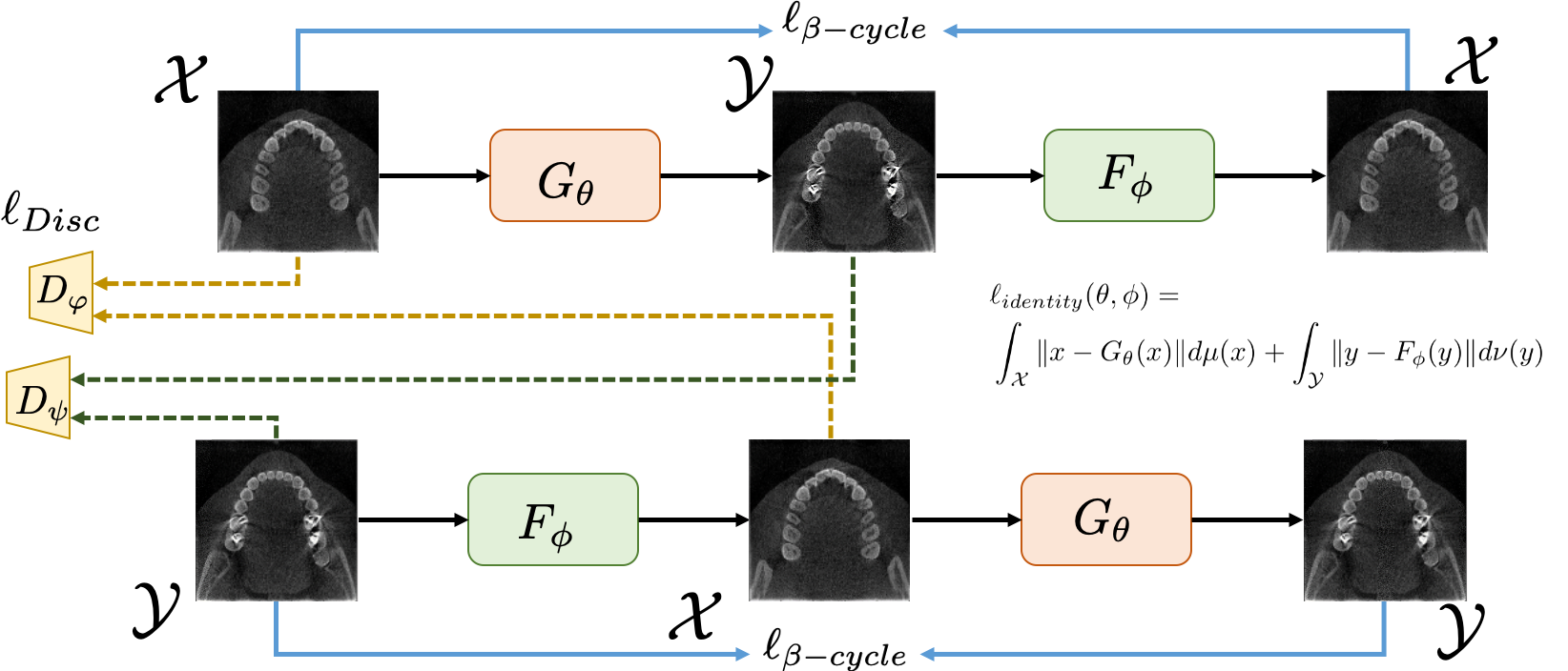}}
\caption{Overall architecture of the proposed neural network approach:  $\Xc$: artifact-free image domain,  $\Yc$: metal artifact image domain.}
\label{fig_architecture_of_network}
\end{figure}

Additionally, in many metal artifact removal problems, the amount of metal artifacts varies and sometimes artifact-free images
could be erroneously used as an input for $G_\theta$. In this case, the same image should be
produced as the output of $G_\theta$. Similarly,  when a metal artifact image is used as an input to $F_\phi$,
the output should be the same image. This can be implemented by an identity loss given by
\begin{align}\label{eq:ourLSnonblind}
&\ell_{identity}(\theta,\phi)=  \notag \\
&\int_\Xc  \|x- G_\theta(x)\|d\mu(x) + \int_\Yc \|y- F_\phi(y)\|d\nu(y) 
\end{align}
Considering all together, our final optimization problem for MAR is given
by
\begin{eqnarray}\label{eq:MARcycleGAN}
\min_{\phi,\theta}\max_{\psi,\varphi}\ell_{MAR}(\theta,\phi;\psi,\varphi)
\end{eqnarray}
where 
\begin{eqnarray}
&\ell_{MAR}(\theta,\phi;\psi,\varphi):= \notag\\
& \lambda \ell_{\beta-cycle}(\theta,\phi) +\ell_{Disc}(\theta,\phi;\psi,\varphi)   + \gamma\ell_{identity}(\theta,\phi)
\end{eqnarray}
where $\gamma>0$ is the hyper-parameter for the identity loss.
The resulting network architecture is illustrated in Fig.~\ref{fig_architecture_of_network}.

\subsection{Geometry of Attention}

In deep neural 
network implementation of attention, two types of attentions are often used: spatial and channel attentions.
To understand these,  a feature map is first  represented by
\begin{align}\label{eq:X}
Z =&\begin{bmatrix} z_1  & \cdots &z_C \end{bmatrix} \in \Rd^{HW\times C},
\end{align}
where 
$z_c  \in \Rd^{HW\times 1}$ refers to the $c$-th column vector of $Z$, which
represents the vectorized feature map of size of 
 $H\times W$  at the $c$-th channel.
 Then, attended feature map $Y \in \Rd^{HW\times C}$ is computed by matrix multiplication:
 \begin{align}\label{eq:attention}
 Y= A Z T
 \end{align}
 where $A \in \Rd^{HW\times HW}$ corresponds to
 a spatial attention map, whereas $T\in \Rd^{C\times C}$
 is a channel attention map.

According to the recent theory of deep convolutional
framelets \cite{ye2018deep},  the expression in \eqref{eq:attention} is exactly the same 
as the $1\times 1$ convolution operation  followed by global
pooling operation.
That said, the main difference of attention module is that
the $1\times 1$  filter kernel and global pooling are estimated
from the feature map rather than pre-trained so that more
data adaptivity can be obtained.

\begin{figure}[!htb]
\centerline{\includegraphics[width=1.0\columnwidth]{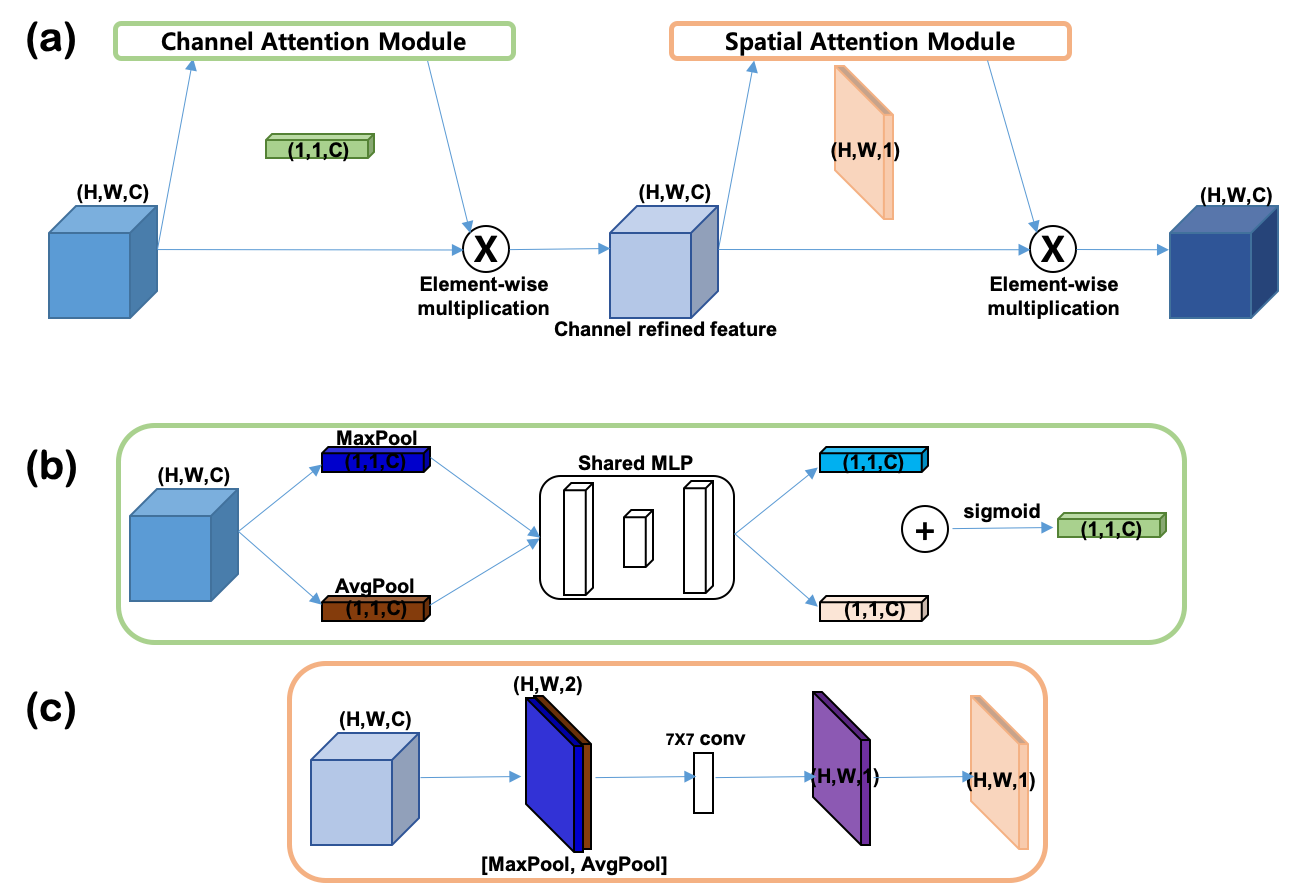}}
\caption{(a) Diagram of CBAM with  (b) channel attention and  (c) spatial attention module.}
\label{fig_cbam}
\end{figure}

In practice, the channel attention map $T$
is implemented as a diagonal matrix
so that each diagonal element represent the weight for each channel.
On the other hand, the spatial attention map $A$ is usually calculated
as a full matrix so that  global information of the features are used
to compute the attended feature map. The main design
criterion is to reduce the computational complexity while maintaining
the feature dependent expressivity.  As such, the convolutional
block attention module (CBAM) \cite{woo2018cbam}  shown in Fig. \ref{fig_cbam} has both
 channel and the spatial attention with relatively small
 computational complexity to achieve the goal.
In the following, we describe CBAM in more details.

\subsubsection{Channel attention module}As each channel of a feature map is considered as a feature detector, channel attention focuses on `what' are important channels given an input image\cite{woo2018cbam}. In order to compute this module efficiently, we squeezed the spatial dimension of the input feature map using both the average pooling for aggregating spatial information, and the max pooling for gathering another important clue about distinctive object features. After that, we passed the two squeezed features through the multi-layer-perceptron (MLP) layer to find 
each channel weighting parameters.

\subsubsection{Spatial attention module}Different from the channel attention, the spatial attention focuses on `where' is an informative part \cite{woo2018cbam}. Spatial attention module also used both the average pooling and the max pooling for memory efficiency. We used the 7$\times$7 convolution operator in order to reflect the spatial domain information. The 7$\times$7 convolution can reflect as wide range of spatial information as it could but not the entire.  

We added CBAM to the skip and the concatenation layers in the generator in order to emphasize a certain part of information when it was delivered from the encoder to the decoder.

\section{Method}
\begin{figure}[!b]
\centering
{\includegraphics[width=0.8\columnwidth]{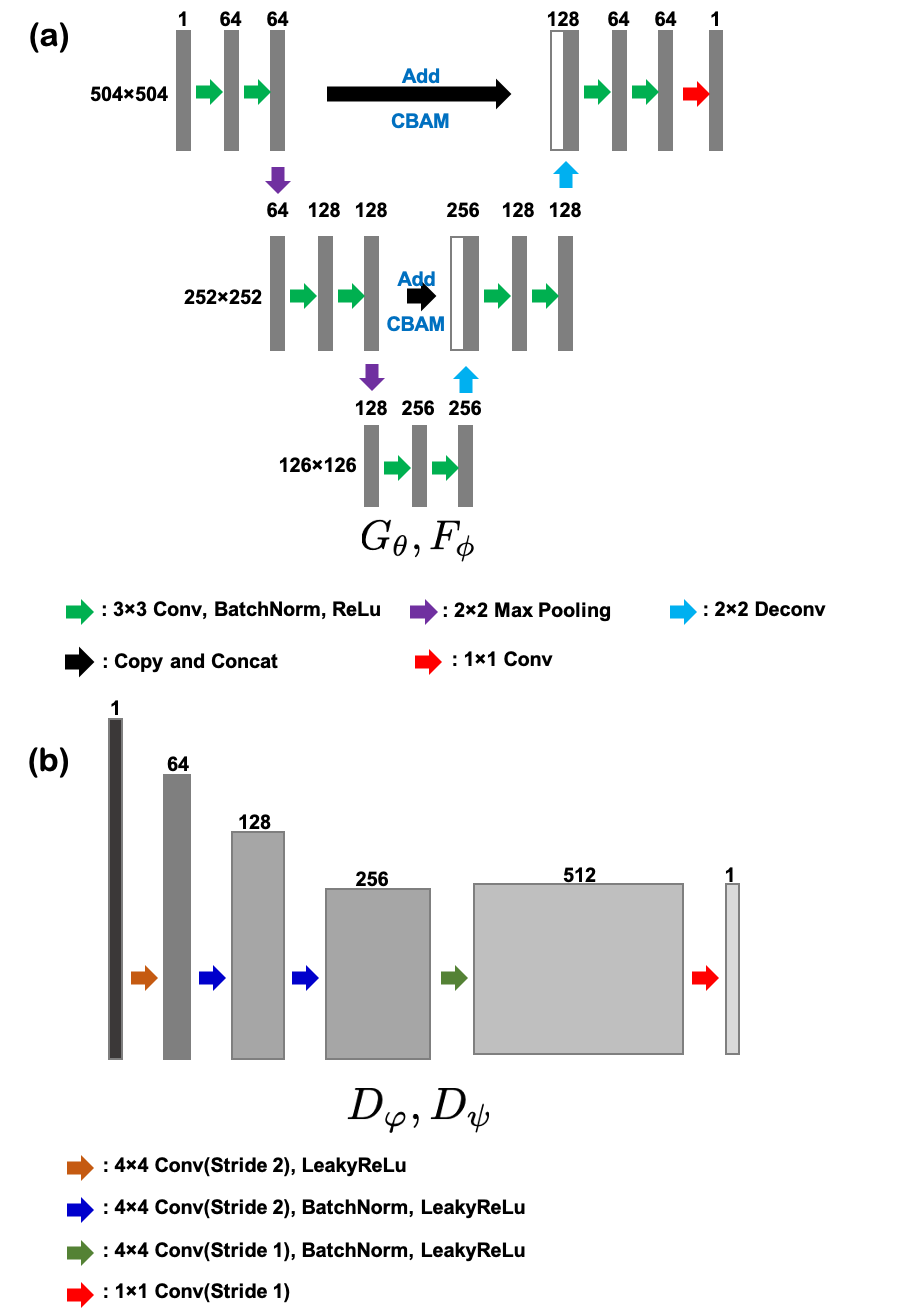}}
\caption{Architecture of  (a) generator and  (b) discriminator.}
\label{fig_model}
\end{figure}
\begin{figure*}[!t]
\centering
{\includegraphics[width=1.6\columnwidth]{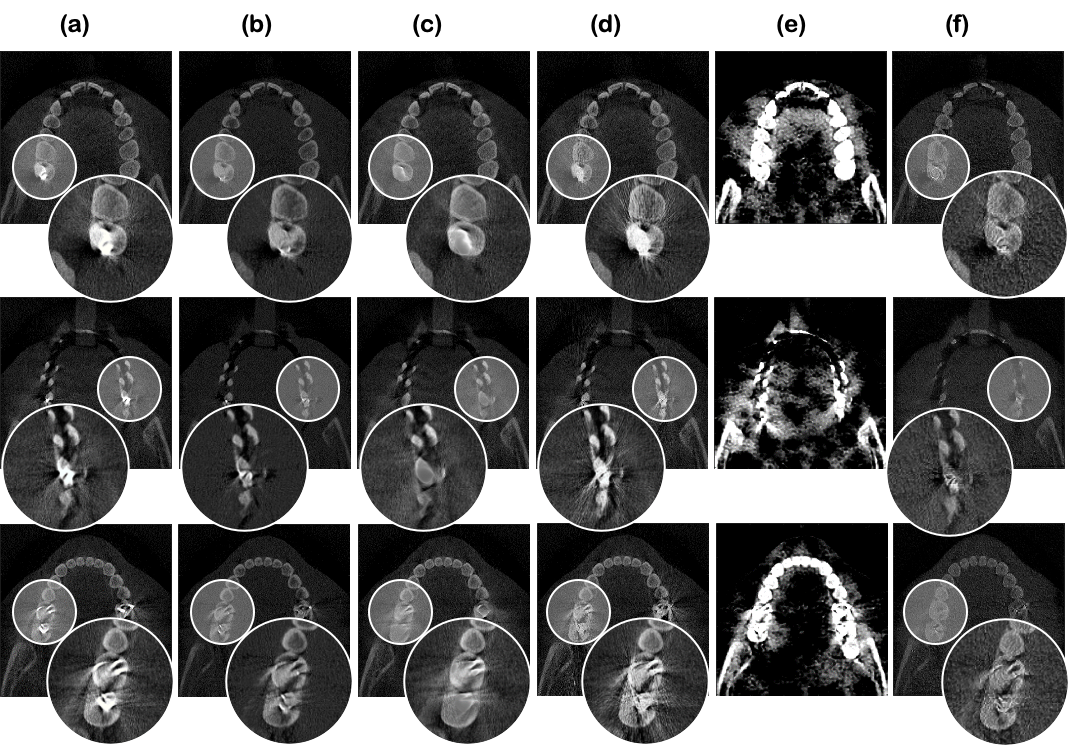}}
\caption{(a) Input images with metal artifacts. Results of the metal artifact removal using  (b) the proposed method, (c) LI, (d) NMAR, (e) ADN with downsampled input, and  (f) the proposed method with downsampled input.}
\label{fig_artifactresult}
\end{figure*}

\subsection{DataSet}
\subsubsection{Real Metal Artifact Data}In this study, we utilized images obtained from a real dental CT scanner from a local vendor.  From the equal-spaced conebeam projection data, we reconstructed the CT images by FDK. The view-angle is from 0$^{\circ}$  to 180$^{\circ}$ . The $x$-$y$ matrix size of the reconstructed images is $504 \times 504$ and the number of  $z$  slices is 400. Out of five patients' data, three patients' data were used as a training set, one patient's data were used as a validation set, and the other patient's data were used for a test set. All images were visually checked to distinguish metal artifact or not. Among 1,200 training slices, 800 slices were with metal artifacts and the remaining 400 slices were without metal artifacts. Because there were no labels, the performance could only be checked qualitatively. 

\subsubsection{Synthetic Metal Artifact Data}The development of MAR algorithms with real samples leads to the difficulty of quantitative evaluation, as there is no clean ground-truth corresponding to a artifact image. Without ground truths, it is not possible to calculate quantitative metrics for the image reconstruction such as the peak signal-to-noise ratio (PSNR) and structural similarity index metric (SSIM)\cite{wang2004image}. For quantitative evaluation of the algorithms, we added synthesized metal artifact to clean data. We randomly selected 10,997 artifact-free CT images from  Liver Tumor Segmentation Challenge (LiTS) dataset\cite{christ2017lits} and followed the method from Convolutional neural network based metal artifact reduction (CNNMAR)\cite{zhang2018convolutional} to synthesize metal artifacts. For making the metal artifacts, we used the code uploaded by the CNNMAR author using the Matlab for a fair comparison.
Specifically, to generate the paired data for training, it simulates the beam hardening effect and Poisson noise during the synthesis of metal-affected polychromatic projection data from artifact-free CT images. 
The number of metal  in a random position is set to be 1 to 2. 

For network training, we used 5,860 images to make synthetic metal artifact data, 4,115 images as clean data. We used 122 images to make synthetic metal artifact data, 192 images at clean data. And we tested 373 images synthesized metal artifact, 335 metal-free images. The size of input image is $256 \times 256$. Before making metal artifact, we downsampled the images using bilinear interpolation, because the full size image ($512 \times 512$) was too big to train the ADN method, whereas the full size image was not a problem with our method. In fact, this is another important advantage of our method.

\subsection{Proposed Network Architecture}
For the generators $G_\theta$ and $F_\phi$ in our MAR model, we used the U-net structure with the attention module in skip and concatenation as shown in Fig. \ref{fig_model}(a). 
A green arrow in Fig. \ref{fig_model} is the basic operator and consists of $3 \times 3$ convolutions followed by a rectified linear unit (ReLU) and batch normalization. The purple arrow is a $2 \times 2$ average pooling operator. And a blue arrow is a $3 \times 3$ deconvolution.  A red arrow is a simple $1 \times1$ convolution operator. In addition a black arrow is a skip and concatenation operator adding CBAM. This attention module consists of two sub-modules, one is channel attention module and the other is spatial attention module. 
The discriminators $D_\varphi$ and $D_\psi$ are constructed based on the structure of PatchGAN\cite{bai2015lidocaine}, which penalizes image patches to capture the texture and style of images. We used PatchGAN composed of four convolution layers and a fully connected layer with the batch normalization.

\begin{figure}[!htb]
\centering
{\includegraphics[width=.9\columnwidth]{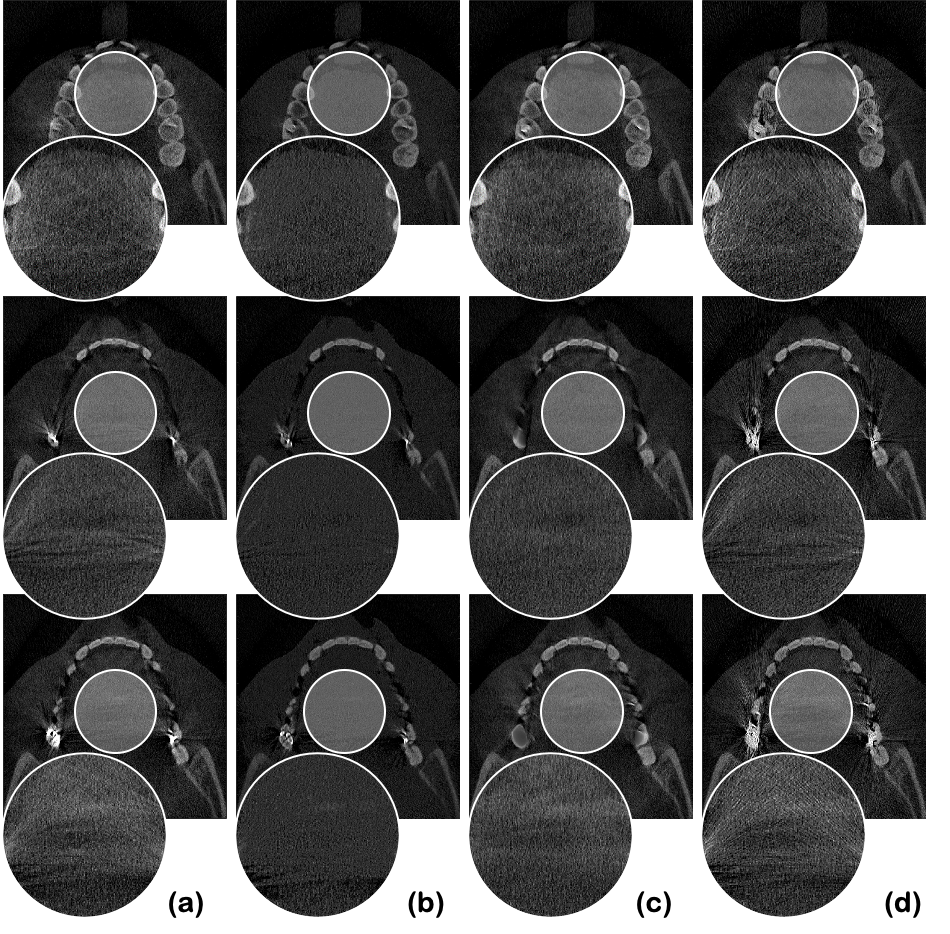}}
\caption{(a) Input images with metal artifacts. Results of the metal artifact removal using  (b) the proposed method, (c) LI, and (d) NMAR. Circles magnify the background image.}
\label{fig_artifactresult_back}
\end{figure}

\begin{figure}[!htb]
\centering
{\includegraphics[width=0.9\columnwidth]{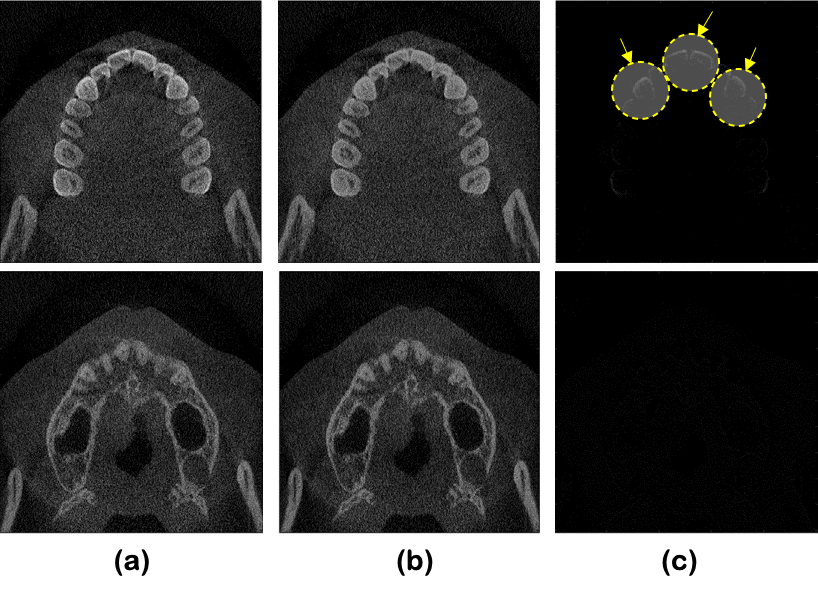}}
\vspace*{-0.5cm}
\caption{Results of the non-metal artifact images processed by the generator $G_\phi$. (a) Input without without metal artifacts,
 (b) network output, and  (c) the difference between input and output.}
\label{fig_nonartifactresult}
\end{figure}

\subsection{Training Details}
\subsubsection{Real Metal Artifact Data}The mini-batch was used as 1 and the size of input image is $504 \times 504$. And the trained network were used for same size images at the inference phase. 
We randomly shuffled images from each metal and non-metal group,  and then we used unmatched data for the training. The network was trained by solving the optimization problem \eqref{eq:MARcycleGAN}  with $\lambda$ = 10,  $\beta$ = 10, and $\gamma$ = 1. 
{Since real metal artifacts are originated from complicated physical phenomenon such as beam-hardening,
photon starvation, etc.,  a large $\beta$ value gives less emphasis on the artifact image generation but more focus
on the artifact-free image generation, which we found useful in real data case. 
}
Moreover, 
as there could be beamhardening artifacts even in the images without metal artifacts, we tried to reduce the identity loss ratio, lessening the value of the hyper-parameter that is involving property that do not need to be changed. 

Adam optimizer was used to optimize the loss function with $\beta_1$ = 0.5 and $\beta_2$ = 0.999. We performed early stopping at 50 epochs, since the early stopping was shown to work as a regularization. The convolution kernels were initialized by xavier initializer. The learning rate was $2 \times 10^{-3}$. We implemented our model using the Tensorflow framework with a NVIDIA GeForce GTX 1080 Ti GPU. 

Further evaluation was conducted after sub-sampling with sampling factor 2 to compare performance with ADN method. Because the full size image was too big to train the ADN method, so we apply ADN after downsampling.

\subsubsection{Synthetic Metal Artifact Data}The mini-batch was used as 1 and the size of input image is $256 \times 256$ that is down-sampling with sampling factor 2.  The downsampling was done for a fair quantitative comparison with ADN which can only work with small size images.
Except for some parameters, we did the same as the real dataset experiment. The network was trained by solving the optimization problem  \eqref{eq:MARcycleGAN} with $\lambda$ = 10,  $\beta$ = 1 and $\gamma$ = 5.  {In the synthetic experiments,
the artifact generation procedure is relatively simple, so we used the same weight on the two statistical distances, i.e. $\beta=1$.}
Additionally, because there is no artifacts in the images with no metal artifacts, we used the larger identity loss ratio in contrast to real dataset experiment.
\section{EXPERIMENTAL RESULTS}
\begin{figure*}[!t]
\centering
{\includegraphics[width=1.6\columnwidth]{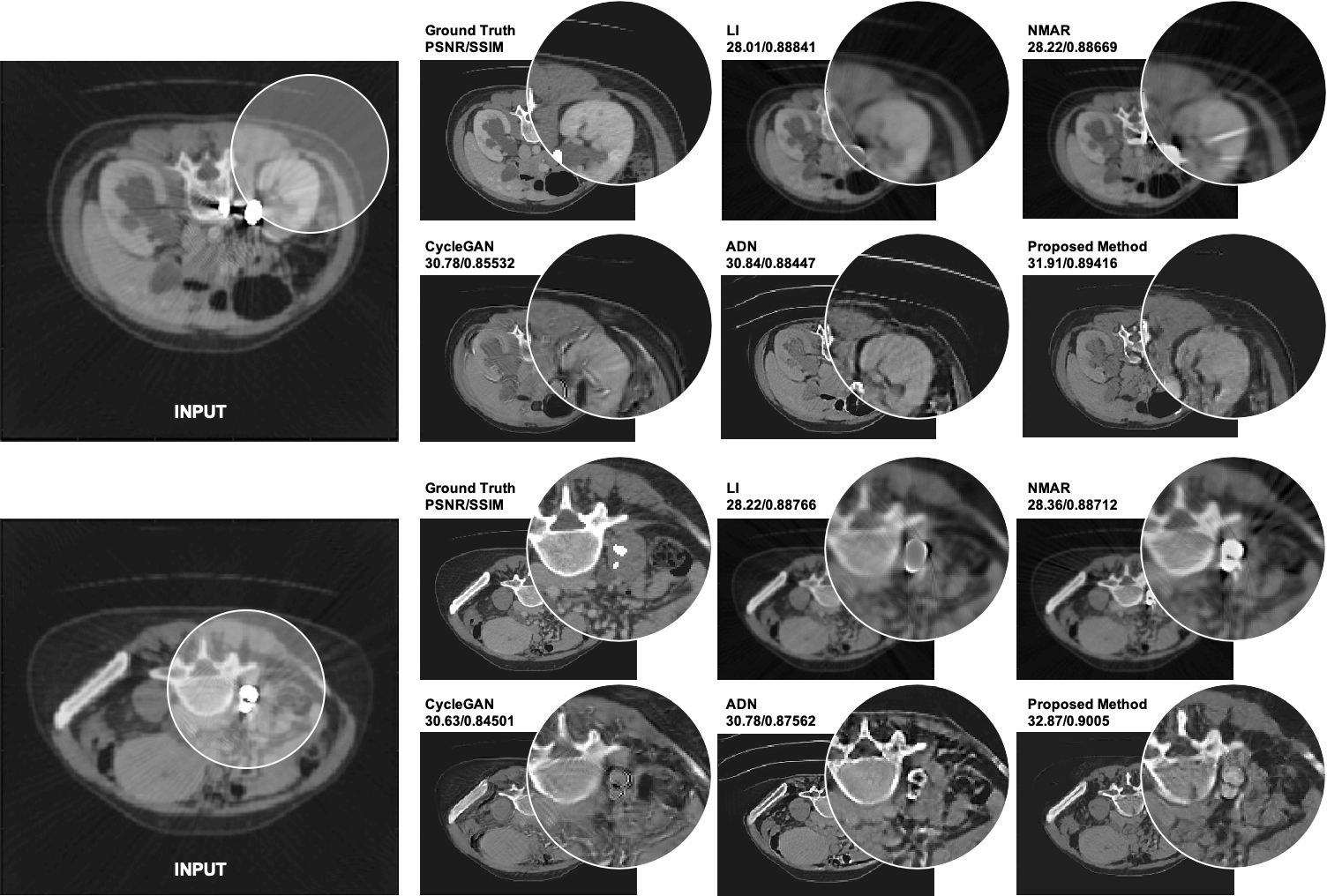}}
\caption{Visual comparisons of various MAR algorithm  results for LiTS CT dataset with synthesized metal artifacts.}
\label{fig_artifact_simulation_result}
\end{figure*}

\begin{figure*}[!t]
\centering
{\includegraphics[width=1.7\columnwidth]{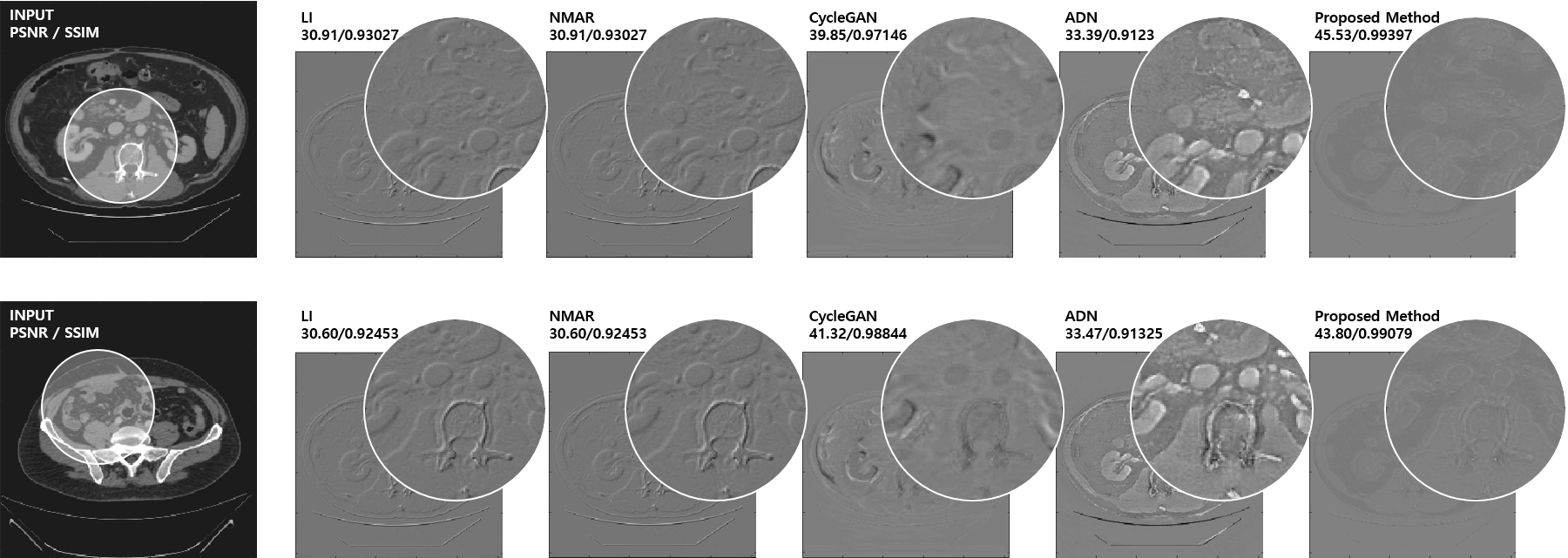}}
\caption{Visual comparisons of various MAR algorithm  results  for the LiTS dataset with no metal artifacts. For accurate verification, the difference between input image and algorithm output were magnified with the same pixel window.}
\label{fig_nonartifact_simulation_result}
\end{figure*}

\begin{figure}[!htb]
\centering
{\includegraphics[width=1.0\columnwidth]{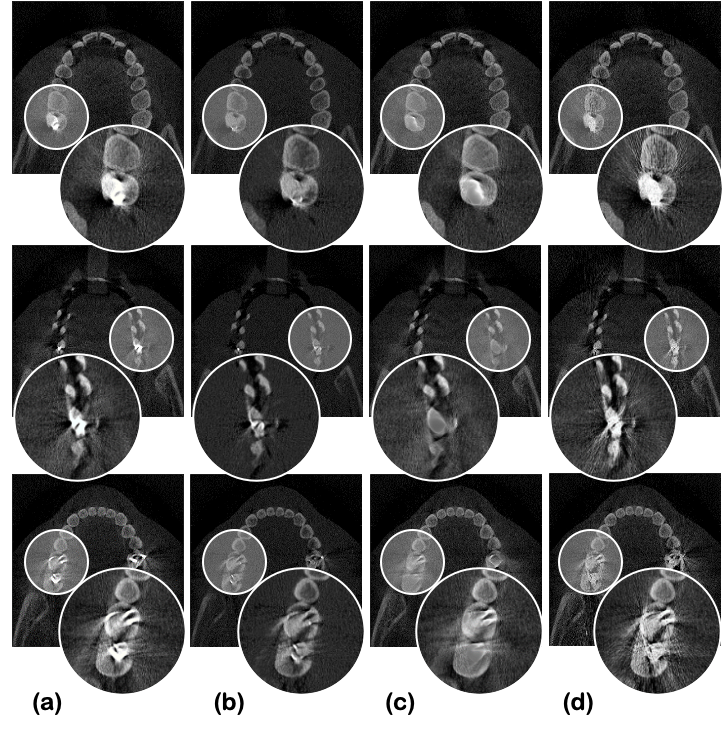}}
\vspace*{-0.5cm}
\caption{Ablation study results: (a) input, (b) the proposed method,  and the ablated network (c) without CBAM module and (d) $\beta=1$.}
\label{fig_ablation_study_result}
\end{figure}

\subsection{Real Experiments}
Fig. \ref{fig_artifactresult} is the result of MAR approaches on the our dental CT dataset by LI \cite{kalender1987reduction}, NMAR\cite{meyer2010normalized}, ADN\cite{liao2019adn}, and our method. As shown in the figure, the proposed method outperform the others. Our method in  Fig. \ref{fig_artifactresult}(b) has successfully removed the metal artifacts with little loss of dental information.  In addition, as shown in Fig. \ref{fig_artifactresult}, the results of proposed method show better quality compare to traditional methods, LI and NMAR. 

We tried to train the ADN method with the original image size for comparison but failed due to a lack of memory, so we reduced the sample size $252 \times 252$.
In spite of this, 
 in the real dental CT images,  ADN which needed artifact images as input did not work well as shown in  Fig. \ref{fig_artifactresult}(e), perhaps because of the beamhardening artifacts on teeth which was not modeled in ADN.  On the other hand,  for both full size image
in  Fig. \ref{fig_artifactresult}(b) and downsampled image in
 Fig. \ref{fig_artifactresult}(f),  our method successfully reduces the metal artifacts.
 
Another important artifacts from metals are banding artifact in the homogeneous tissue regions.
Fig. \ref{fig_artifactresult_back} shows that the shading patterns resulted from the metal artifacts has been reduced by
our methods, whereas other methods failed.

We also checked how it works for images without metal artifacts. If there is no metal artifact, there shall be no difference between input and output.
Fig. \ref{fig_nonartifactresult} shows that images without metal artifacts are successfully recovered by our method. Moreover, as shown in Fig. \ref{fig_nonartifactresult}(a), there was an additional positive effect that the generator have improved the quality of images removing beam-hardening artifacts.

\subsection{Synthetic Metal Artifact Experiments}
For the quantitative evaluation, we performed inferences using synthetic metal artifact data and calculated quantitative metrics such as PSNR and SSIM. Fig. \ref{fig_artifact_simulation_result} is the result of MAR approaches on the LiTS dataset with synthesized metal artifact by LI, NMAR, CycleGAN, ADN and our method. As shown in the figure, our method outperform the others in terms of PSNR and SSIM for all images. Furthermore, as shown in Table \ref{tab_quantitative_LiTS_synthesized}, we observed that the mean PSNR and SSIM values of  our results from 373 test images are highest among other methods, which confirms that our method improves the performance. We also checked how it works for images without metal artifacts. If there is no metal artifact, there shall be no difference between input and output. As shown from the difference images in  Fig. \ref{fig_nonartifact_simulation_result}, it can be confirmed that there is little difference between output and input when we use the proposed method,
whereas other methods produce significant errors. For  accurate  verification, the difference between input image and algorithm output was  magnified in their pixel window. Also, as shown in the Table \ref{tab_quantitative_LiTS}, we observed that the mean PSNR and SSIM values of  our results from 335 test images are highest among other methods. This is clearly the difference when compared with ADN.

\begin{table} [!h]
\centering
\caption{Quantitative evaluation of MAR approaches on the LiTS dataset with synthesized metal artifact}
\setlength{\tabcolsep}{15pt}
\renewcommand{\arraystretch}{1.2}
\begin{tabular}{c  c  c  c}\hline
& Algorithm & PSNR & SSIM\\\hline
Unsupervised  & CycleGAN & 29.29 & 0.8477 \\
& ADN & 29.95 & 0.8879 \\
& Proposed & \textbf{30.78} & \textbf{0.8912} \\\hline
Conventional& LI & 27.47 & 0.8852 \\
& NMAR & 27.68 & 0.8807 \\\hline
\end{tabular}
\label{tab_quantitative_LiTS_synthesized}
\end{table}
\begin{table} [!h]
\centering
\caption{Quantitative evaluation of MAR approaches on the LiTS dataset with no metal artifacts}
\setlength{\tabcolsep}{15pt}
\renewcommand{\arraystretch}{1.2}
\begin{tabular}{c  c  c  c}\hline
& Algorithm & PSNR & SSIM\\\hline
Unsupervised  & CycleGAN & 41.11 & 0.9834 \\
& ADN & 33.85 & 0.9173 \\
& Proposed & \textbf{44.74} & \textbf{0.9922} \\\hline
Conventional& LI & 30.73 & 0.9295 \\
& NMAR & 30.73 & 0.8801 \\\hline
\end{tabular}
\label{tab_quantitative_LiTS}
\end{table}

\subsection{Ablation Study}
\subsubsection{Result of CycleGAN without CBAM}We also compared the results with respect to the existence of CBAM in Fig. \ref{fig_ablation_study_result} (b), (c). Fig. \ref{fig_ablation_study_result}(b) was obtained by using the proposed method, and Fig. \ref{fig_ablation_study_result}(c) was obtained by using a generator without CBAM module. The hyper-parameter setting for both was identical. After the training without CBAM, we found that the artifacts in the background had become fainter compared to those of the input. However, compared to the result with the proposed method, the degree of removal of metal artifacts had decreased. We also observed that the real metallic objects are incorrectly
removed.   Therefore, we found that CBAM is an essential part of our MAR.  

\subsubsection{Dependency on the disentanglement parameter $\beta$}In Fig. \ref{fig_ablation_study_result} (b) and (d), we could compare the results according to the value of the hyper-parameter. While Fig. \ref{fig_ablation_study_result}(b) was obtained using the proposed method with $\beta=10$, Fig. \ref{fig_ablation_study_result}(d) was obtained using the standard setting for CycleGAN ($\beta=1$). The generator with CBAM was used in both. In Fig. \ref{fig_ablation_study_result} (d), we found that the metal artifact removal  in the background was
not as effect as in (b) by the proposed method.  By increasing $\beta$ to focus more on the artifact-free images,
we could disentangle the metal-artifact generation and  solve the MAR more effectively.

\section{Conclusion}

In this paper, we proposed a novel $\beta$-cycleGAN with an attention module for the metal artifact removal in  CT data. To alleviate the problem of the unpaired data in practice, we trained the MAR network in an unsupervised manner. Although there were several successful unsupervised
deep learning approaches, it was difficult to apply them to our MAR problem since their networks were too heavy. 
Because metal artifacts occur local position with globally radiating artifact, we used CBAM  to focus on important features in both spatial
and channel domain. Moreover, 
we introduced a disentanglement parameter $\beta$ that imposes
relative importance on the reconstructed artifact-free images compared to the artifact generation process. This proves to be effective for real data
where the generation of metal artifacts is very complicated due to beam hardening, photon starvations, etc.
Our attention-guide $\beta$-cycleGAN network could be trained efficiently in an unsupervised manner, and effectively mitigated
the metal artifacts in both synthetic and real data.  Moreover,  when the algorithm was applied to non-artifact images, the images were rarely damaged, which demonstrated its robustness to other algorithms.

\appendix

The derivation of the dual formula is simple modification of the technique in \cite{sim2019optimal}.
We first define the optimal joint measure $\pi^*$ for the primal problem \eqref{eq:betaprimal}. 
Using the Kantorovich dual formulations, we have the following two equalities:
\begin{align}
K:=& \int_{\Xc\times \Yc} c(x,y) d\pi^*(x,y)\notag \\
=  \max_{\varphi} & \left\{ \int_\Yc \inf_x \{ c(x,y) -D_\varphi(x) \}d\nu(y)  + \int_\Xc D_\varphi(x) d\mu(x)  \right\} \label{eq:e1}\\ 
=\max_{\psi} &\left\{ \int_\Xc \inf_y \{c(x,y) -D_\psi(y) \}d\mu(x) + \int_\Yc \psi 
(y)d\nu(y)\right\} \label{eq:e2}
\end{align}
where
$ c(x,y):=\frac{1}{\beta}\|y- F_\phi(x)\|+\| G_\Theta(y) - x\|.$
Using  $1/\beta$-Lipschitz continuity of the Kantorovich potentials, we have
\begin{align*}
&-D_\varphi(G_\Theta(y))\leq \| G_\Theta(y) - x\| -D_\varphi(x) \leq  c(x,y)-D_\varphi(x) \\
&-D_\psi(F_\phi(x))  \leq  \frac{1}{\beta}\|y-F_\phi(x)\| -D_\psi(y) \leq  c(x,y) -D_\psi(y) 
\end{align*}
This leads to two lower bounds and by taking the average of the two, we have
\begin{align*}
K\geq \frac{1}{2}\ell_{Disc}(\theta,\phi;\varphi,\psi)
\end{align*}
where $\ell_{Disc}$  is defined in \eqref{eq:disc}.
For and upper bound,  instead of finding the $\inf_x$, we choose {$x=G_\Theta(y)$} in \eqref{eq:e1}; similarly, instead of $\inf_y$,  we chose $y=F_\phi(x)$
in \eqref{eq:e2}. By taking the average of the two upper bounds, we have
\begin{eqnarray*}
K&\leq  &  \frac{1}{2} \left\{\max_{\varphi}\int_\Xc D_\varphi(x) d\mu(x)-\int_\Yc D_\varphi(G_\Theta(y))d\nu(y) \right.\\
&&\quad \left. +\int_\Yc  \|y- F_\phi(G_\theta(y))\| d\nu(y)\right. \\
& &\quad+\max_{\psi}\int_{\Yc} D_\psi(y)  d\nu(y) - \int_\Xc D_\psi(F_\phi(x))  d\mu(x)  \\
&& \quad+\left. 
\int_\Xc\| G_\Theta(F_\phi(x)) - x\| d\mu(x)
\right\}\\
&=& \frac{1}{2}\left\{\ell_{Disc}(\theta,\phi;\varphi,\psi)+\ell_{\beta-cycle}(\theta,\phi)\right\}
\end{eqnarray*}
where   $\ell_{\beta-cycle}$  is defined \eqref{eq:betacycle}.
%
%
The remaining part of the proof for the dual formula \eqref{eq:Dd} is a simple repetition of the techniques  in \cite{sim2019optimal}.

\bibliographystyle{IEEEtran}
\bibliography{MAR_unsupervised_cyclegan_cbam_bib}

\begin{thebibliography}{10}
\providecommand{\url}[1]{#1}
\csname url@samestyle\endcsname
\providecommand{\newblock}{\relax}
\providecommand{\bibinfo}[2]{#2}
\providecommand{\BIBentrySTDinterwordspacing}{\spaceskip=0pt\relax}
\providecommand{\BIBentryALTinterwordstretchfactor}{4}
\providecommand{\BIBentryALTinterwordspacing}{\spaceskip=\fontdimen2\font plus
\BIBentryALTinterwordstretchfactor\fontdimen3\font minus
  \fontdimen4\font\relax}
\providecommand{\BIBforeignlanguage}[2]{{%
\expandafter\ifx\csname l@#1\endcsname\relax
\typeout{** WARNING: IEEEtran.bst: No hyphenation pattern has been}%
\typeout{** loaded for the language `#1'. Using the pattern for}%
\typeout{** the default language instead.}%
\else
\language=\csname l@#1\endcsname
\fi
#2}}
\providecommand{\BIBdecl}{\relax}
\BIBdecl

\bibitem{feldkamp1984practical}
L.~A. Feldkamp, L.~C. Davis, and J.~W. Kress, ``Practical cone-beam
  algorithm,'' \emph{Josa a}, vol.~1, no.~6, pp. 612--619, 1984.

\bibitem{tomography2003principles}
C.~Tomography, ``Principles, design, artifacts, and recent advances,''
  \emph{Jiang Hsieh}, 2003.

\bibitem{zhao2000x}
S.~Zhao, D.~Robeltson, G.~Wang, B.~Whiting, and K.~T. Bae, ``X-ray ct metal
  artifact reduction using wavelets: an application for imaging total hip
  prostheses,'' \emph{IEEE transactions on medical imaging}, vol.~19, no.~12,
  pp. 1238--1247, 2000.

\bibitem{kalender1987reduction}
W.~A. Kalender, R.~Hebel, and J.~Ebersberger, ``Reduction of ct artifacts
  caused by metallic implants.'' \emph{Radiology}, vol. 164, no.~2, pp.
  576--577, 1987.

\bibitem{mahnken2003new}
A.~H. Mahnken, R.~Raupach, J.~E. Wildberger, B.~Jung, N.~Heussen, T.~G. Flohr,
  R.~W. G{\"u}nther, and S.~Schaller, ``A new algorithm for metal artifact
  reduction in computed tomography: in vitro and in vivo evaluation after total
  hip replacement,'' \emph{Investigative radiology}, vol.~38, no.~12, pp.
  769--775, 2003.

\bibitem{meyer2010normalized}
E.~Meyer, R.~Raupach, M.~Lell, B.~Schmidt, and M.~Kachelrie{\ss}, ``Normalized
  metal artifact reduction (nmar) in computed tomography,'' \emph{Medical
  physics}, vol.~37, no.~10, pp. 5482--5493, 2010.

\bibitem{shepp1982maximum}
L.~A. Shepp and Y.~Vardi, ``Maximum likelihood reconstruction for emission
  tomography,'' \emph{IEEE transactions on medical imaging}, vol.~1, no.~2, pp.
  113--122, 1982.

\bibitem{wang1996iterative}
G.~Wang, D.~L. Snyder, J.~A. O'Sullivan, and M.~W. Vannier, ``Iterative
  deblurring for ct metal artifact reduction,'' \emph{IEEE transactions on
  medical imaging}, vol.~15, no.~5, pp. 657--664, 1996.

\bibitem{de2001iterative}
B.~De~Man, J.~Nuyts, P.~Dupont, G.~Marchal, and P.~Suetens, ``An iterative
  maximum-likelihood polychromatic algorithm for ct,'' \emph{IEEE transactions
  on medical imaging}, vol.~20, no.~10, pp. 999--1008, 2001.

\bibitem{wang2018conditional}
J.~Wang, Y.~Zhao, J.~H. Noble, and B.~M. Dawant, ``Conditional generative
  adversarial networks for metal artifact reduction in ct images of the ear,''
  in \emph{International Conference on Medical Image Computing and
  Computer-Assisted Intervention}.\hskip 1em plus 0.5em minus 0.4em\relax
  Springer, 2018, pp. 3--11.

\bibitem{zhang2018convolutional}
Y.~Zhang and H.~Yu, ``Convolutional neural network based metal artifact
  reduction in x-ray computed tomography,'' \emph{IEEE transactions on medical
  imaging}, vol.~37, no.~6, pp. 1370--1381, 2018.

\bibitem{lin2019dudonet}
W.-A. Lin, H.~Liao, C.~Peng, X.~Sun, J.~Zhang, J.~Luo, R.~Chellappa, and S.~K.
  Zhou, ``Dudonet: Dual domain network for ct metal artifact reduction,'' in
  \emph{Proceedings of the IEEE Conference on Computer Vision and Pattern
  Recognition}, 2019, pp. 10\,512--10\,521.

\bibitem{goodfellow2014generative}
I.~Goodfellow, J.~Pouget-Abadie, M.~Mirza, B.~Xu, D.~Warde-Farley, S.~Ozair,
  A.~Courville, and Y.~Bengio, ``Generative adversarial nets,'' in
  \emph{Advances in neural information processing systems}, 2014, pp.
  2672--2680.

\bibitem{zhu2017unpaired}
J.-Y. Zhu, T.~Park, P.~Isola, and A.~A. Efros, ``Unpaired image-to-image
  translation using cycle-consistent adversarial networks,'' in
  \emph{Proceedings of the IEEE international conference on computer vision},
  2017, pp. 2223--2232.

\bibitem{villani2008optimal}
C.~Villani, \emph{Optimal transport: old and new}.\hskip 1em plus 0.5em minus
  0.4em\relax Springer Science \& Business Media, 2008, vol. 338.

\bibitem{peyre2019computational}
G.~Peyr{\'e}, M.~Cuturi \emph{et~al.}, ``Computational optimal transport,''
  \emph{Foundations and Trends{\textregistered} in Machine Learning}, vol.~11,
  no. 5-6, pp. 355--607, 2019.

\bibitem{sim2019optimal}
B.~Sim, G.~Oh, S.~Lim, and J.~C. Ye, ``Optimal transport, cycle{GAN}, and
  penalized {LS} for unsupervised learning in inverse problems,'' \emph{arXiv
  preprint arXiv:1909.12116}, 2019.

\bibitem{larochellesupplementary}
H.~Larochelle and G.~E. Hinton, ``Learning to combine foveal glimpses with a
  third-order boltzmann machine,'' in \emph{Advances in neural information
  processing systems}, 2010, pp. 1243--1251.

\bibitem{itti1998model}
L.~Itti, C.~Koch, and E.~Niebur, ``A model of saliency-based visual attention
  for rapid scene analysis,'' \emph{IEEE Transactions on pattern analysis and
  machine intelligence}, vol.~20, no.~11, pp. 1254--1259, 1998.

\bibitem{rensink2000dynamic}
R.~A. Rensink, ``The dynamic representation of scenes,'' \emph{Visual
  cognition}, vol.~7, no. 1-3, pp. 17--42, 2000.

\bibitem{corbetta2002control}
M.~Corbetta and G.~L. Shulman, ``Control of goal-directed and stimulus-driven
  attention in the brain,'' \emph{Nature reviews neuroscience}, vol.~3, no.~3,
  pp. 201--215, 2002.

\bibitem{woo2018cbam}
S.~Woo, J.~Park, J.-Y. Lee, and I.~So~Kweon, ``Cbam: Convolutional block
  attention module,'' in \emph{Proceedings of the European Conference on
  Computer Vision (ECCV)}, 2018, pp. 3--19.

\bibitem{higgins2017beta}
I.~Higgins, L.~Matthey, A.~Pal, C.~Burgess, X.~Glorot, M.~Botvinick,
  S.~Mohamed, and A.~Lerchner, ``$\beta$-{VAE}: Learning basic visual concepts
  with a constrained variational framework.'' \emph{Iclr}, vol.~2, no.~5, p.~6,
  2017.

\bibitem{de2000reduction}
B.~De~Man, J.~Nuyts, P.~Dupont, G.~Marchal, and P.~Suetens, ``Reduction of
  metal streak artifacts in x-ray computed tomography using a transmission
  maximum a posteriori algorithm,'' \emph{IEEE transactions on nuclear
  science}, vol.~47, no.~3, pp. 977--981, 2000.

\bibitem{liao2019adn}
H.~Liao, W.-A. Lin, S.~K. Zhou, and J.~Luo, ``Adn: Artifact disentanglement
  network for unsupervised metal artifact reduction,'' \emph{IEEE Transactions
  on Medical Imaging}, 2019.

\bibitem{ranzini2020combining}
M.~Ranzini, I.~Groothuis, K.~Kl{\"a}ser, M.~J. Cardoso, J.~Henckel,
  S.~Ourselin, A.~Hart, and M.~Modat, ``Combining multimodal information for
  metal artefact reduction: An unsupervised deep learning framework,''
  \emph{arXiv preprint arXiv:2004.09321}, 2020.

\bibitem{odena2017conditional}
A.~Odena, C.~Olah, and J.~Shlens, ``Conditional image synthesis with auxiliary
  classifier gans,'' in \emph{Proceedings of the 34th International Conference
  on Machine Learning-Volume 70}.\hskip 1em plus 0.5em minus 0.4em\relax JMLR.
  org, 2017, pp. 2642--2651.

\bibitem{lucas2018mixed}
T.~Lucas, C.~Tallec, J.~Verbeek, and Y.~Ollivier, ``Mixed batches and symmetric
  discriminators for gan training,'' \emph{arXiv preprint arXiv:1806.07185},
  2018.

\bibitem{zhang2018self}
H.~Zhang, I.~Goodfellow, D.~Metaxas, and A.~Odena, ``Self-attention generative
  adversarial networks,'' \emph{arXiv preprint arXiv:1805.08318}, 2018.

\bibitem{kingma2013auto}
D.~P. Kingma and M.~Welling, ``Auto-encoding variational {B}ayes,'' \emph{arXiv
  preprint arXiv:1312.6114}, 2013.

\bibitem{kullback1997information}
S.~Kullback, \emph{Information theory and statistics}.\hskip 1em plus 0.5em
  minus 0.4em\relax Courier Corporation, 1997.

\bibitem{kang2018deep}
E.~Kang, W.~Chang, J.~Yoo, and J.~C. Ye, ``Deep convolutional framelet denosing
  for low-dose ct via wavelet residual network,'' \emph{IEEE transactions on
  medical imaging}, vol.~37, no.~6, pp. 1358--1369, 2018.

\bibitem{song2020unsupervised}
J.~Song, J.-H. Jeong, D.-S. Park, H.-H. Kim, D.-C. Seo, and J.~C. Ye,
  ``Unsupervised denoising for satellite imagery using wavelet subband
  cyclegan,'' \emph{arXiv preprint arXiv:2002.09847}, 2020.

\bibitem{arjovsky2017wasserstein}
M.~Arjovsky, S.~Chintala, and L.~Bottou, ``Wasserstein {GAN},'' \emph{arXiv
  preprint arXiv:1701.07875}, 2017.

\bibitem{lim2019cyclegan}
S.~Lim, S.-E. Lee, S.~Chang, and J.~C. Ye, ``Cyclegan with a blur kernel for
  deconvolution microscopy: Optimal transport geometry,'' \emph{IEEE Trans. on
  Computational Imaging (in press), also available as arXiv preprint
  arXiv:1908.09414}, 2020.

\bibitem{mao2017least}
X.~Mao, Q.~Li, H.~Xie, R.~Y. Lau, Z.~Wang, and S.~Paul~Smolley, ``Least squares
  generative adversarial networks,'' in \emph{Proceedings of the IEEE
  international conference on computer vision}, 2017, pp. 2794--2802.

\bibitem{gulrajani2017improved}
I.~Gulrajani, F.~Ahmed, M.~Arjovsky, V.~Dumoulin, and A.~C. Courville,
  ``Improved training of {W}asserstein {GAN}s,'' in \emph{Advances in neural
  information processing systems}, 2017, pp. 5767--5777.

\bibitem{ye2018deep}
J.~C. Ye, Y.~Han, and E.~Cha, ``{Deep convolutional framelets: A general deep
  learning framework for inverse problems},'' \emph{SIAM Journal on Imaging
  Sciences}, vol.~11, no.~2, pp. 991--1048, 2018.

\bibitem{wang2004image}
Z.~Wang, A.~C. Bovik, H.~R. Sheikh, and E.~P. Simoncelli, ``Image quality
  assessment: from error visibility to structural similarity,'' \emph{IEEE
  transactions on image processing}, vol.~13, no.~4, pp. 600--612, 2004.

\bibitem{christ2017lits}
P.~Bilic, P.~F. Christ, E.~Vorontsov, G.~Chlebus, H.~Chen, Q.~Dou, C.-W. Fu,
  X.~Han, P.-A. Heng, J.~Hesser \emph{et~al.}, ``The liver tumor segmentation
  benchmark ({LiTS}),'' \emph{arXiv preprint arXiv:1901.04056}, 2019.

\bibitem{bai2015lidocaine}
Y.~Bai, T.~Miller, M.~Tan, L.~S.-C. Law, and T.~J. Gan, ``Lidocaine patch for
  acute pain management: a meta-analysis of prospective controlled trials,''
  \emph{Current medical research and opinion}, vol.~31, no.~3, pp. 575--581,
  2015.

\end{thebibliography}


\end{document}